\documentclass[twocolumn]{aastex631} 
\usepackage{csvsimple}
\usepackage{multirow,bigdelim}

\usepackage{amsmath}
\usepackage{mathtools}
\usepackage{hyperref}

\newcommand{\T}{{\rm T}}

\usepackage{colortbl}
\definecolor{tablegray}{rgb}{0.89, 0.89, 0.89}
\shorttitle{ERES 2023: Early-Career Conference Organization}
\shortauthors{The ERES 2023 Organizing Committee}

\begin{document}

\title{Emerging Researchers in Exoplanetary Science (ERES):\\ Lessons Learned in Conference Organization for Early-Career Researchers}
\correspondingauthor{W. Garrett Levine}

\email{garrett.levine@yale.edu}

%\author{The ERES 2023 Organizing Committee}

\author[0000-0002-1422-4430]{W. Garrett Levine}
\affil{Dept. of Astronomy, Yale University. New Haven, CT 06511, USA}

\author[0000-0002-4836-1310]{Konstantin Gerbig}
\affil{Dept. of Astronomy, Yale University. New Haven, CT 06511, USA}

\author[0000-0003-3179-5320]{Emma M. Louden}
\affil{Dept. of Astronomy, Yale University. New Haven, CT 06511, USA}

\author[0000-0003-0834-8645]{Tiger Lu}
\affil{Dept. of Astronomy, Yale University. New Haven, CT 06511, USA}

\author[0000-0003-2803-6358]{Cheng-Han Hsieh}
\affil{Dept. of Astronomy, Yale University. New Haven, CT 06511, USA}

\author[0000-0003-3987-3776]{Christopher O'Connor}
\affil{Dept. of Astronomy and Cornell Center for Astrophysics \& Planetary Science, Cornell University, Ithaca, NY 14853, USA}

\author[0000-0001-9222-4367]{Rixin Li}
\affil{Dept. of Astronomy, University of California, Berkeley, Berkeley, CA 94720, USA}

\author[0000-0002-3610-6953]{Jiayin Dong}
\affil{Center for Computational Astrophysics, Flatiron Institute, New York, NY 10010, USA}

\begin{abstract}
Since 2015, the Emerging Researchers in Exoplanetary Science (ERES) conference has provided a venue for early-career researchers in exoplanetary astronomy, astrophysics, and planetary science to share their research, network, and build new collaborations. ERES stands out in that it is spearheaded by early-career researchers, providing a unique attendance experience for the participants and a professional experience for the organizers. In this Bulletin, we share experiences and lessons learned from the perspective of the organizing committee for the 2023 edition of ERES. For this eighth ERES conference, we hosted over 100 participants in New Haven, CT, for a three-day program. This manuscript is aimed primarily toward groups of early-career scientists who are planning a conference for their fields of study. We anticipate that this Bulletin will continue dialogue within the academic community about best practices for equitable event organization.
\vspace{10mm}
\end{abstract}

% Add official keywords
\keywords{}

\section{Introduction} \label{sec:intro}

Attending conferences and presenting one's research is an integral part of building an academic career, especially for early-career researchers such as undergraduates, graduate students, and postdoctoral fellows. For those who eventually pursue careers outside of the academic track, the public speaking and networking skills developed at these meetings are equally critical. To that end, the Emerging Researchers in Exoplanetary Science (ERES) symposium was founded in 2015 by graduate students at Cornell, Penn State, and Yale. While the first iterations were only open to scientists at those host institutions for financial and logistical reasons, ERES has opened attendance to participants from outside of the Northeast. In recent years, ERES has grown into a preeminent global venue for early-career exoplanetary scientists to share their research and network with others at similar career stages; the 2022 and 2023 iterations hosted by Penn State and Yale attracted 85 and 106 attendees, respectively.

Notably, ERES organization has always been led by early-career researchers. While this structure introduces some challenges -- ERES is often the first time that most organizing committee (OC) members put together a conference -- the benefits of organically growing the event within the early-career community are numerous. First, the scientific content, discussion topics, and networking events are better tailored to the attendees' needs. Second, the experience equips OC members with logistical skills that will pay dividends in their careers. Third, the peer-led OC promotes an atmosphere of camaraderie towards an inclusive, relatable conference, allowing attendees to engage as embodied participants.

Results from a post-conference survey from ERES 2023 were overwhelmingly positive (see Sec. \ref{sec:conference}), affirming the value of a space designed for early-career researchers to present and network. Through a generous grant from the Heising-Simons Foundation, the conference had no registration fee, provided meals and lodging for all participants, and covered travel costs for those who would otherwise be unable to attend.

The ERES 2023 OC consisted exclusively of volunteer early-career researchers: six graduate students and two postdoctoral fellows. In this Bulletin, we provide a case study on conference organization for the astronomy community from an early-career perspective. In other disciplines, early-career astrobiologists have been particularly prolific at publicly documenting their efforts toward organizing events \citep{abGradCon2021, astrobioEarlyCareer2022}. Within astronomy, the Future of Meetings group \citep{moss2021betterNormal} has discussed how conference organization has changed due to the COVID-19 pandemic. We anticipate that this manuscript will add to a growing body of literature and benefit future groups of early-career researchers who organize their own symposia.

Our OC benefited from numerous online\footnote{\url{https://www.exordo.com/blog/organising-conference-tips/}} resources\footnote{\url{https://www.london.ac.uk/venues/blog/how-plan-academic-conference}} on conference organization, and we refer the reader to these references for comprehensive, step-by-step checklists. Here, we focus on what we believe were the most consequential choices that we made in preparing for ERES 2023. To this end, Section \ref{sec:conference} provides an overview of the event. Then, Section \ref{sec:preconference} highlights pre-conference logistical preparation. Section \ref{sec:scienceOC} details abstract sorting and other scientific program considerations. Next, Section \ref{sec:midconference} describes logistics during the conference itself. Finally, Sections \ref{sec:postconference} and \ref{sec:conclusions} serve to summarize lessons learned and to look ahead to future ERES symposia.

\section{ERES 2023 Symposium} \label{sec:conference}

The eighth ERES was hosted by Yale University from June 18-20, 2023. All conference-related events, including the scientific program, group meals, and lodging, were held at the Omni Hotel at Yale in downtown New Haven, CT. Some attendees arrived on Sunday, June 18, but participants from closer to New Haven arrived on the morning of Monday, June 19. In-person attendance was the only option for ERES 2023. The 106 attendees hailed from 53 institutions and 4 countries.

\begin{figure*}
\centering
\includegraphics[width=0.83\linewidth]{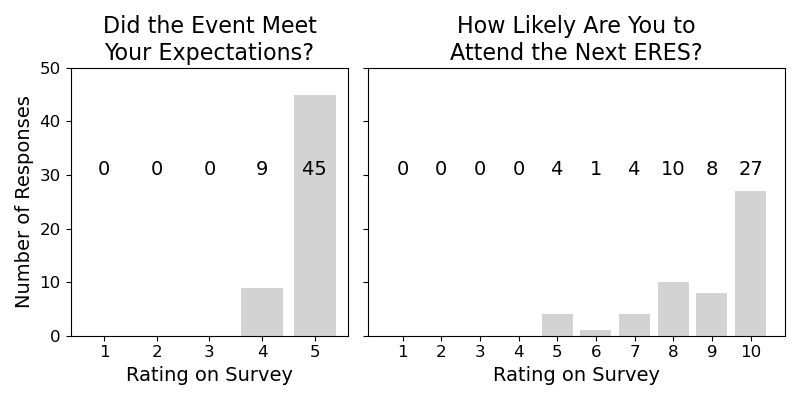}
\caption{Results from a backward-looking and a forward-looking question from the ERES 2023 post-conference survey. On both questions, higher numerical values correspond to more favorable scores. This survey was optional and sent to participants via the attendee Slack workspace after the conference ended.}
\label{fig:nextYear}
\end{figure*}

The block schedule for ERES 2023 can be found in Appendix \ref{sec:schedule}. The scientific program featured 37 oral presentations and 64 posters. The presenters were a mix of undergraduates, graduate students, and postdoctoral researchers. Oral sessions were in plenary format, with one room dedicated to these talks. All posters were on display for the entire conference, and poster sessions were held jointly with coffee breaks between oral sessions.

In addition, a professional development panel with two senior astronomers covered topics such as building a professional network, academic and non-academic career paths, job application strategies, interviewing and negotiation techniques, and managing teams in academia, industry, and government. After predetermined questions, there was open Q\&A.

As a condition of registering for the conference, all attendees and organizers agreed to follow a Code of Conduct; the full text is available in Appendix \ref{sec:CoCText}. Some of the ERES Code of Conduct's content used language borrowed from the existing AAS Code of Ethics\footnote{\url{https://aas.org/policies/ethics}}. During the conference, designated OC members were available via in-person conversation, phone, and email in case of concerns related to the Code. An anonymous form was also available as an alternative reporting avenue.

Outside of the science program, buffet-style meals were provided for all participants to encourage networking and to reduce the cost of attendance. On the evening of Monday, June 19, a trivia night (with small prizes for the winners) was offered as an optional informal social event. Since ERES was not held on Yale's campus itself, a sense of place was established by welcoming a visit from Yale's bulldog mascot, Handsome Dan, for lunch and group photos on Tuesday, June 20. 

Especially for early-career researchers, conferences are an opportunity to assess the host institution as a potential future employer. To facilitate this outcome, we offered an optional one-hour campus tour after the scientific program had concluded. This event was not on the original agenda and was added by popular demand.

After the conference, attendees were asked to complete an online feedback form. These results indicate that ERES 2023 was successful in facilitating networking, scientific discussion, and community-building among the participants. Figure \ref{fig:nextYear} provides a summary from two questions, and Appendix \ref{sec:survey} provides additional data.

\section{Pre-Conference Logistics} \label{sec:preconference}

Organizing ERES was a team effort, including all members of the OC, the Yale Astronomy department and business office, the participants, and the Heising-Simons Foundation. Our overarching philosophy was to make decisions that simplified the planning process and logistics. Buy-in, resilience, and planning in advance were critical to the success of ERES. We established timelines in Fall 2022 for when we needed to reach milestones, giving us a clear roadmap (see Appendix \ref{sec:Gantt}).

\subsection{Organizing Committee Structure}

Two Yale graduate students had served on the OC for ERES 2022 at Penn State, which smoothed the transition to the ERES 2023 organization. In addition, the lead organizer from Penn State during the previous year served on the ERES 2023 OC in an advisory role. This continuity was critical to ensuring the success of ERES 2023. This systematic handover provided a strong foundation to capitalize on past experience, and the tradition of mentorship within the ERES OC underscores the essence of the conference itself.

In total, the ERES 2023 committee consisted of five graduate students from Yale Astronomy (ERES 2023 host), one former graduate student from Penn State Astronomy (ERES 2022 host), and one graduate student and one postdoc from Cornell Astronomy (likely ERES 2024 host). Cornell OC members secured departmental support for tentatively hosting ERES 2024, ensuring that their experience would benefit the next ERES.

Having an OC composed of early-career researchers was empowering. We strongly recommend this format to other groups of early-career researchers who wish to organize their own symposia. Our faculty mentors were unwaveringly supportive and provided insightful advice, but early-career researchers were the ultimate stakeholders and the ones most affected by the conference's purpose, outcome, and long-term viability. Our approach solidified the genuine impact and relevance of ERES to the early-career exoplanet community.

\subsection{OC Time Management}

We deliberately minimized the time involved for all OC members. All personnel were only required to attend meetings relevant to their responsibilities, but updates were sent on a shared Slack workspace for all to see. In February 2023, we held an in-person meeting with the entire OC. Otherwise, all catch-ups involved only a subset of the group. Since planning for the logistical aspects and the scientific program often did not overlap, we divided members into a logistical organizing committee and a scientific organizing committee early in the process. The leaders of these groups managed and delegated tasks among OC members with availability.

Building buffers into the planning timeline and having redundancy on the OC was important to accommodate the ebbs and flows of early-stage academic careers. Nobody was a professional conference organizer and could not prioritize ERES over research. A mutual understanding that we were all in similar positions helped to cultivate a strong team. OCs of early-career researchers should be large in size, with members participating as their schedules allow. 

\subsection{Budget Management}

Conference participants were nearly all undergraduates, post-baccalaureate researchers, graduate students, or postdocs. As such, we attempted to eliminate barriers to attendance that may affect early-career scientists. Many of these hurdles are financial in nature, and ERES has been fortunate to be supported by the Heising-Simons Foundation since 2018. With a grant for ERES 2023, participants without other travel funding could attend at no cost. Moreover, this funding from Heising-Simons covered lodging and food for all participants. While group meals are not included at many astronomy conferences, we believe that removing this financial barrier and adding time for informal networking was beneficial in the early-career conference setting.

One barrier to participation that we were unable to resolve was the prerequisite in-person attendance. Travel added logistical complications, especially for those doing summer programs and for researchers located far from New Haven. Whether we should offer a hybrid option was discussed at length amongst the OC, but we decided that our budget was best used to improve the in-person experience and to defray costs for participants traveling to New Haven. We believe that the community-building goal of ERES is only possible in-person. Even recording the talks for asynchronous viewing would have come at a substantial cost and logistical burden.

ERES 2023 had record interest, as over 160 early-career researchers initially registered. Attrition was an issue later -- we discuss this matter further in Section \ref{sec:postconference}. We did not expect this number of sign-ups and subsequently modified our initial plan to accommodate this interest. The major changes were as follows:

\begin{itemize}
    \item Participants with less than a 4-hour drive to New Haven were asked to arrive on Monday morning instead of Sunday evening, saving costs on Sunday night for hotel rooms and meals.
    \item Those same participants were not provided lodging on Tuesday night, but were provided meals on Monday and Tuesday with all other attendees.
    \item Instead of a traditional 9AM-5PM schedule on both days, we began the scientific program at 10:45 AM on Monday morning to give Monday morning arrivals more time to get to New Haven.
    \item We asked for volunteers to share hotel rooms, to which we got an outstanding response. Originally, we had budgeted for all participants to have single rooms. With the buy-in from attendees, we were able to use only the originally anticipated number of hotel rooms. We understood that not everybody is comfortable sharing rooms, so we emphasized that this request was entirely voluntary. 
\end{itemize}

To explore whether we could increase our total budget, we initiated discussions with philanthropic divisions of corporations in the space industry. We were unable to secure additional funding from these sources, but we were generously granted backstop funding from the Yale Astronomy and Physics departments to accommodate the additional interest. Because of the aforementioned attrition, we did not draw heavily on these funds. In total, we were able to provide a no-cost conference to attendees because of the participants' shared sense of responsibility toward the mission of ERES.

\subsection{Promotion \& Attendee Communication}

ERES maintains an account on X (formerly Twitter) to publicize the conference. We posted when registration opened and directed people to a standalone website (made with Canva) for more information. As of October 2023, the home page has over 2800 views, and the registration posts on X received over 10,000 impressions. 

\begin{figure}[!h]
\includegraphics[width=1.00\linewidth]{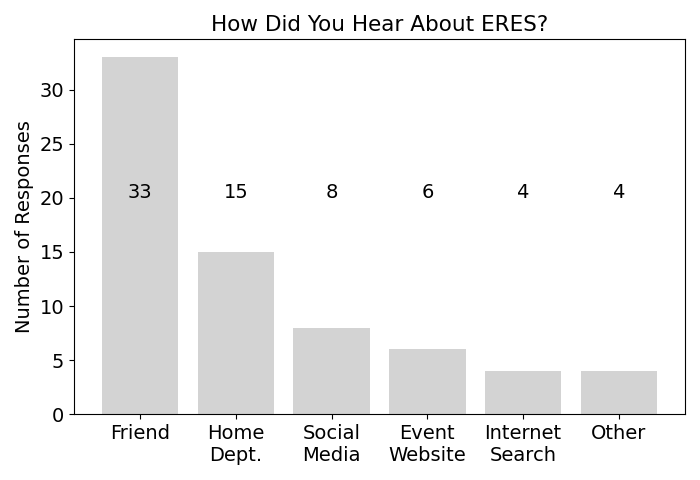}
\caption{Results from a question on the post-conference survey asking how attendees learned about the conference. Word-of-mouth was the predominant component, although department mailing lists were also important.}
\label{fig:hearAbout}
\end{figure}

Despite these efforts on social media, a plurality of participants who responded to the post-conference survey reported hearing about ERES primarily via word-of-mouth (see Figure \ref{fig:hearAbout}). While this result is a testament to the grassroots nature of ERES, it also indicates that our social media presence could expand. For example, future organizers could build a standing website that is passed from one OC to the next one. In addition to providing a permanent online home for ERES, the website could link to a standing Zenodo repository where conference abstracts can be assigned a DOI. Future OCs could also create accounts on BlueSky or other X alternatives.

To register for the ERES, participants were required to join a Slack workspace (free version) and were told that all future communication from the OC would be on this platform. Limiting messages to Slack helped the OC to streamline communication. Having a channel dedicated to logistical questions helped all OC members stay updated on all communication with attendees. Email was the primary means of communication in past years, but we are pleased with our decision to use Slack instead.

During registration, we asked attendees to submit information that would be necessary for reimbursement. This framework reduced the amount of tracking-down that needed to be done after the conference. At sign-up time, we also communicated what documentation of travel costs would be required for reimbursement. By setting a reimbursement processing plan with the Yale Astronomy business office before sending abstract acceptance emails, we reduced the overall logistical burden.

We required that attendees book travel within two weeks of receiving their acceptance to the conference to enable a relatively accurate headcount for hotel rooms and catering. In retrospect, we should have only required that airfare be booked -- our request led to many questions about booking ground transportation.

Between registration and the conference, we maintained frequent communication with attendees regarding travel and lodging. Participants were divided among the OC members such that all attendees had an assigned point of contact. This decision spread out the workload and allowed for more personalized communication. We maintained a list of FAQs on the website but should have kept a running log of questions from last year's ERES to be proactive. Recommendations for travel to New Haven were among the most common questions.

\section{Scientific Program Organization} \label{sec:scienceOC}

\begin{figure*}
    \centering
    \includegraphics[width = \linewidth]{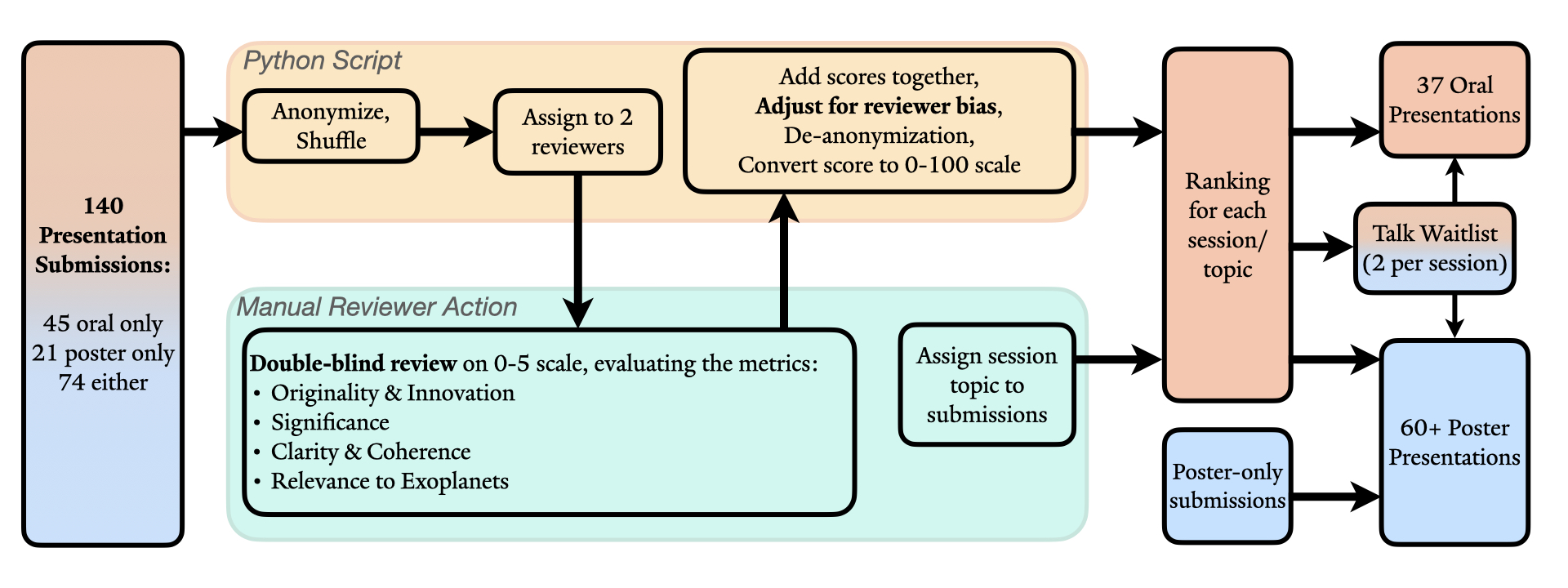}
    \caption{Flowchart visualizing the selection process for oral presentations, centered around a double-blind review with the aim to minimize biases. Notably, the review process did not distinguish between submissions that requested only an oral presentation versus those interested in either talks or posters. Every submission was accepted for a poster presentation; hence, the discrepancy between the number of initial submissions and final presentations is due to attrition as discussed in Sect.~\ref{sec:attrition}.}
    \label{fig:soc_flowchart}
\end{figure*}

The ERES 2023 abstract submission form asked participants to indicate their interest in an oral presentation, a poster, or both. We originally planned to have 32 talks, but we received 45 ``oral only" and 74 ``either oral or poster" submissions (versus 21 ``poster only"). Even after expanding to 37 talks, these slots were 2x oversubscribed due to the record number of registrants.

Here, we detail the double-blind review process that we implemented to sort the abstracts and share other aspects of the scientific program organization that we believe could benefit future early-career conferences.

\subsection{Abstract Review Process}

For reference, our procedure for reviewing submitted abstracts is illustrated in Figure~\ref{fig:soc_flowchart}. All information from the registration form, including titles, abstracts, and talk versus poster preferences, was saved in a Google Sheet. We wrote a \texttt{python} script that pulled abstracts of participants who requested talks from this sheet, anonymized the entries, and randomly shuffled the order. This code is available upon reasonable request to the authors. Each abstract was distributed to two random members of the science OC for a blind review. Importantly, each submission was reduced to only the title and abstract; the career stage was omitted. Reviewers flagged conflicts of interest if they recognized a submission's author, in which case the abstract was redistributed randomly to another person. Scores were assigned on a 0-5 scale on each of the following metrics: Originality \& Innovation, Significance, Clarity \& Coherence, and Relevance to ERES. Each submission thus received eight scores from two reviewers for each metric, with a maximum possible score of 40.

Once the scoring was concluded, we normalized scores such that each reviewer's collection of submissions had an average score of 2.5 on each individual metric. This procedure compensated for different mean scores from each reviewer and assumed that the samples of submissions each reviewer is assigned to are statistically identical in quality. We believe that this assumption was largely realized since each reviewer had approximately 35 randomly assigned abstracts. Summed scores from all categories from both reviewers were projected onto a 1-100 scale. Then, de-anonymization led to the final ranking of all abstract submissions. 

Ideally, each abstract would have been reviewed by all reviewers. This task was infeasible, given the number of submissions and the time constraints of OC members. In addition, the performed correction implicitly assumed that each reviewer's scores followed the same underlying distribution (as we solely corrected for the mean). This was demonstrably not the case in the actual review, as some used the full 0-5 scales, some gave similar scores for all submissions, and some reviewers even produced bimodal distributions. We ultimately decided against performing additional adjustments, such as re-scaling scores by the standard deviation of the respective reviewer, to avoid unnecessary complexity on what would have been a small effect on the relative rankings.

In total, our abstract review methodology generated a ranking that was broadly representative of the attendees' home institutions, participant career stages, and preferred pronouns. This result underscores the benefits of a double-blind review. Nonetheless, the ranking did prefer certain sub-fields within the broader exoplanet discipline over others. This outcome was likely indicative of the OC's own research interests; most members were theorists studying celestial dynamics and magnetohydrodynamics. To avoid propagating this bias to the talk schedule, we decided on the science session topics before the double-blind review based on the content of submitted abstracts and filled the available slots by category from the ranked subset of abstracts in that group.

Overall, we deemed the abstract review process successful in our goal of reducing implicit bias. One logistical lesson learned is that the abstract submission form should ask registrants to tag their presentation among a group of standardized topics. Asking participants to pre-sort would have saved time for the OC and possibly reduced additional bias, as we had to manually assign topics to all submissions. To foster community engagement and to provide a larger set of reviewers for each abstract, we recommend that future organizers consider including non-OC members (perhaps the conference participants themselves) in the double-blind review.

\subsection{Presentation Types \& Coordination}

We did not anticipate such demand for oral presentations, as we aimed to provide substantial opportunities for poster presenters and talk-givers to share their science. Our goal was to make both formats equally desirable, so the actual balance of submissions was not the intended outcome. We opted for a large space at the hotel for the poster hall that allowed us to display all posters for the whole conference. By allotting four poster sessions, we hoped that all attendees could share their work without feeling rushed. Future groups of early-career conference organizers could set expectations in advance by advertising tentative talk lengths and the amount of time dedicated to the posters during the sign-up period.

ERES has never had keynote presentations, and this year was no exception. Differentiating between early-career researchers would have complicated the organization and introduced additional subjectivity into the abstract sorting process. We are happy with this decision and believe that a standardized talk timeslot promoted a more egalitarian conference culture.

In the week before ERES, oral presenters were asked to send their slides to the science OC chair at least two days before the start of the conference. Edits were allowed until the day before the presentations. Only PowerPoint, Keynote, and PDF presentations were accepted to minimize the possibility of technical problems. In the future, a Google Drive or other repository may streamline this process and reduce the workload for the OC. 

We did not offer onsite printing services for poster presenters, so attendees needed to bring posters to New Haven or find a local printing service. The OC fielded many questions on poster logistics; future organizers could consider providing information on poster printing on the conference website's FAQ section. Notably, the cost of printing an academic poster is a financial barrier to attendance that we did not defray for participants.

\subsection{Career Panel Preparation}

During the past few ERES conferences, the only participation by senior astronomers has been on a professional development panel. At ERES 2022, the panel's focus was the 2020 Decadal Survey and different career pathways in Astronomy. We focused this year on navigating the job market, a relevant topic for early-career attendees. To make the panel helpful for a broad swath of participants, we recruited speakers who pursued various career trajectories after PhD programs in astronomy and astrophysics. We aimed to recruit a professor at an R1 university, a NASA scientist specializing in mission design, and somebody who pursued a career outside of astrophysics. Although we successfully recruited panelists from among the professional networks of OC members to fill the first two categories, the third one went unfilled since no potential panelists were available during the dates of ERES.

For the first part of the panel with predetermined questions, the OC developed discussion topics in consultation with the panelists. Both panelists had similar amounts of time to speak about their backgrounds, and we tried to craft questions to which both panelists could respond. This part of the panel took 40 minutes, leaving the final 20 minutes for open Q\&A from the audience.

The feedback about the career panel on the post-conference survey was overall positive (see Figure \ref{fig:multipanel}), and we recommend that other conferences for early-career researchers build time into the scientific program for similar conversations. Scheduling the scientific program after determining availability for potential panelists may help with finding panelists to participate. ERES is a unique space for conversations about career planning, the state of the discipline, and professional development; panels can serve as structured environments to focus these important discussions.

\section{Conference Execution} \label{sec:midconference}

We knew that establishing a sense of professionalism at the symposium was important, so we aimed to provide a smooth logistical experience and promote productive scientific discussions. All conference-related events were held at the hotel, and we strongly recommend that future events for early-career researchers consider this option. This decision increased accessibility and streamlined the logistics; the scientific program and group meals were located on the same floor, and participants could go to-and-from their rooms during the day. Through a single contact at the Omni, we coordinated the space, lodging, catering, and audiovisual setup.

\subsection{Interfacing with Participants}

Depending on their arrival time, participants could check in either at dinner on Sunday or at breakfast on Monday. We provided nametags, ERES-themed notebooks, stickers, pens, and a Yale Astronomy pizza cutter since New Haven is famous for pizza. We opted for high-quality notebooks, a decision which we believe helped set the tone for the conference as high-quality itself.

We began a pilot program this year where oral session chairs were not solely OC members. We solicited volunteers from the participants themselves, adding to the sense of early-career ownership. For each session, we aimed to have two chairs; sessions before time-sensitive events were co-chaired by OC members. We asked chairs to meet with the session presenters 20 minutes beforehand for an audiovisual check and to ensure that all names and affiliations were pronounced correctly.

At the conclusion of each oral session, a member of the OC spoke to give logistical announcements and to provide participants with directions to the next events on the schedule. These frequent updates contributed to keeping the scientific program on time, which ensured that all oral and poster presenters were given their allotted time. Having a smoothly running program also helped attendees with tight travel plans participate in the entire conference.

During the group lunch on Tuesday, representatives from the Yale Astronomy business office held ``reimbursement office hours," where participants could ask questions about required documentation and discuss edge cases that were not covered by the directions sent to the whole group. By getting ahead of issues that might arise after ERES, we reduced the post-conference burden on both the OC and the business office.

\subsection{OC Assignments and Duties}

Most OC members elected not to contribute an abstract. Having everybody available for logistics during the entire meeting was worthwhile, and we were happy that we did not have to balance preparing for a scientific presentation with organizational responsibilities.

The OC had an ``office," which was a small meeting room next to the oral session hall. Having this space was more important than we had initially realized. This room was a secure place to store materials and personal items, and having a quiet space to take five-minute breaks was helpful for restoring mental stamina. A private area was also helpful for on-the-fly coordination. We recommend that organizers secure a similar space at future conferences.

Two OC roles that kept the conference running smoothly were assigned point people for the catering and audiovisual. These people ensured that meals were ready on-time and that the projector and microphone were ready for each session. The audiovisual lead stayed in the presentation room during all oral sessions and doubled as a photographer to document the event. We later shared the photos with participants, many of whom reported using these pictures on their social media and professional websites. Delegating responsibilities in advance reduced decision fatigue for the OC and provided third-party vendors with a consistent point-of-contact. 

Much of the intra-OC communication during ERES was via Slack. Each night, however, we held a 30-minute synchronous check-in meeting to debrief on the day's activities and to confirm that all tasks were assigned for the next day. This OC meeting served as an important transition from one day to the next.

\section{Post-Conference: Lessons Learned} \label{sec:postconference}

We believe that ERES achieved its goal of providing a venue for scientific discussion and networking among early career researchers, but some aspects would have benefitted from hindsight. Here, we comment on a few areas for improvement that were identified through the post-conference survey and debrief OC discussions.

\subsection{Gathering Community Feedback}

Participation in the post-conference survey was lower than we had hoped, with only approximately 40\% of attendees responding. While a significant fraction of the attendees, we would prefer for this number to approach 100\%. Future ERES conferences could run a feedback survey during the final session to boost participation.

In hindsight, we realize that a gathering of around 100 early-career researchers provides a serendipitous opportunity to gather data on the general health of the community. Opt-in focus groups could ask about challenges faced by the attendees and solicit ideas for early-career researchers to support each other. This initiative would align with ERES' goal of being a positive force towards community-building among the stakeholders.

\subsection{Conference Venue Layout}

Feedback on the post-conference survey indicated that one of the weaknesses of ERES 2023 was the venue layout (see Appendix \ref{sec:survey}). Attendees sitting towards the back of the oral session hall had difficulty seeing the slides, and there was limited access to natural daylight in the conference venue. While OC members did multiple walk-throughs of the venue with hotel staff before the conference, we did not test lines-of-sight for the full conference setup by sitting in various places in the room. An elevated screen or multiple synchronized projectors would have alleviated some of these concerns.

\subsection{Attendee Attrition}
\label{sec:attrition}

Our initial round of registration surpassed 160 sign-ups. As the conference approached, however, attrition was an issue; more than 140 people accepted their conference invitations once abstracts were sorted, but only 106 ended up attending the conference. Many did not cancel until the week before ERES. Because participants could attend ERES at no cost, we suspect that there was less of a psychological and financial barrier to canceling versus conferences that charge registration fees. Logistically, this 35\% dropout rate was a major issue. We budgeted conservatively and assumed that all participants who were signed up at any point in time would actually attend the conference. This approach ensured that we would retain a balanced budget but also meant that our funds were not allocated as efficiently as possible.

We are conflicted on how to account for attrition and reduce this rate. Since we could not overextend funding promises to attendees, we believe that our budgeting approach was warranted. Assuming an attrition rate in our initial budget would have worked for this conference in retrospect, but it was a risk that we deemed not worthwhile. Next year, we will aim to convey the scope of the attrition issue to potential participants while sign-ups are open. The participants are the ultimate stakeholders in ERES, and conveying the same camaraderie during the registration period that was present during the conference itself could help to reduce attrition.

At a structural level, we discussed adding a small registration deposit, which would be refunded after attending ERES. However, this fee would have imposed a financial burden on the attendees and a logistical burden on the organizers. For these reasons, we opted to avoid this path. We leave the topic of handling attrition open for now, but we hope that the above discussion can provide some insight for future conference organizers.

\section{Conclusions \& Looking Ahead} \label{sec:conclusions}

In the past eight years, ERES has grown from a local meeting of a few institutions into a conference of international importance for early-career exoplanet researchers. We are proud of the self-organized nature of ERES and intend to maintain this structure, but we may consider future collaborations with professional societies and new host schools. At present, attendees from outside of the East Coast have needed to travel for several hours to attend ERES every year.

Reflecting on the planning and execution of ERES, we believe that the following are our biggest takeaways:

\begin{itemize}
    \item The attributes required for successfully organizing a conference overlap substantially with those required for success in academic research and in industry careers: time and project management, resilience, and teamwork.
    \item Especially for conferences aimed at early-career researchers, ensuring accessibility is critical. ERES has been successful due to the financial generosity of the Heising-Simons Foundation, financial and administrative support from host departments, and volunteer OCs. Together, these forces have eliminated the cost of attendance for participants without other sources of funding.
    \item Continuity between events in the series is important to retain institutional knowledge, plan effectively, and organize efficiently. Early-career researchers balance numerous academic and professional responsibilities, so mentorship from former OC members and faculty advisors is key.
    \item Buy-in from attendees is a prerequisite for a professional, welcoming, and inclusive event.
\end{itemize}

The engagement, feedback, and progress that we witnessed this year serve as a testament to ERES' value. Conversations with participants suggest that ERES has led to new research collaborations among early-career scientists. With each iteration, we aim to further enrich the community of early-career exoplanet researchers. We are excited about the future trajectory of this event.

Finally, we invite other groups of (early-career or otherwise) researchers who are considering organizing a conference to contact us with any questions about our experience. We would be happy to share our perspective and receive community input to further improve future iterations of ERES. We hope that this Bulletin inspires others to create and lead their own symposia or other events that are tailored toward early-career researchers in their fields. We believe that initiatives like ERES help to advance research and to promote an inclusive community in which to develop future scientific leaders. By coming together as an early-career researcher community, we elevate our own scientific knowledge, broaden our cultural horizons, and ignite the collective curiosity that drives community-driven scientific discovery.

\vspace{1cm}

ACKNOWLEDGEMENTS: We are grateful for the generous financial support from the Heising-Simons Foundation towards the 2018, 2019, 2022, and 2023 iterations of this conference. In addition, we are indebted to the members of past ERES organizing committees for their hard work and dedication. We thank all of the participants from past ERES conferences for their scientific contributions, energy, enthusiasm, and feedback.

We thank Yale Astronomy for their institutional support of ERES and are especially appreciative of the faculty advisors: Priyamvada Natarajan, Greg Laughlin, and Sarbani Basu. Finally, we thank the Yale Astronomy business office -- Hannah Carroll, Maria Troiano, Robyn Lisone, Teena Griggs, Stacey Hampton, Shawna Rodriguez, and Geriana Van Atta -- for going above-and-beyond their departmental responsibilities to help with the event. Without this departmental and administrative support, ERES would not have been possible.

Sections of this manuscript were edited for typographical errors, clarity, and conciseness with assistance from OpenAI's GPT-4. Some figures were made with \texttt{numpy} \citep{harris2020numpy} and \texttt{matplotlib} \citep{hunter2007matplotlib}.

\bibliography{bibliography}
\bibliographystyle{aasjournal}

\appendix

\section{Conference Schedule} \label{sec:schedule}

\begin{figure}[!h]
\centering
\includegraphics[width=1\linewidth]{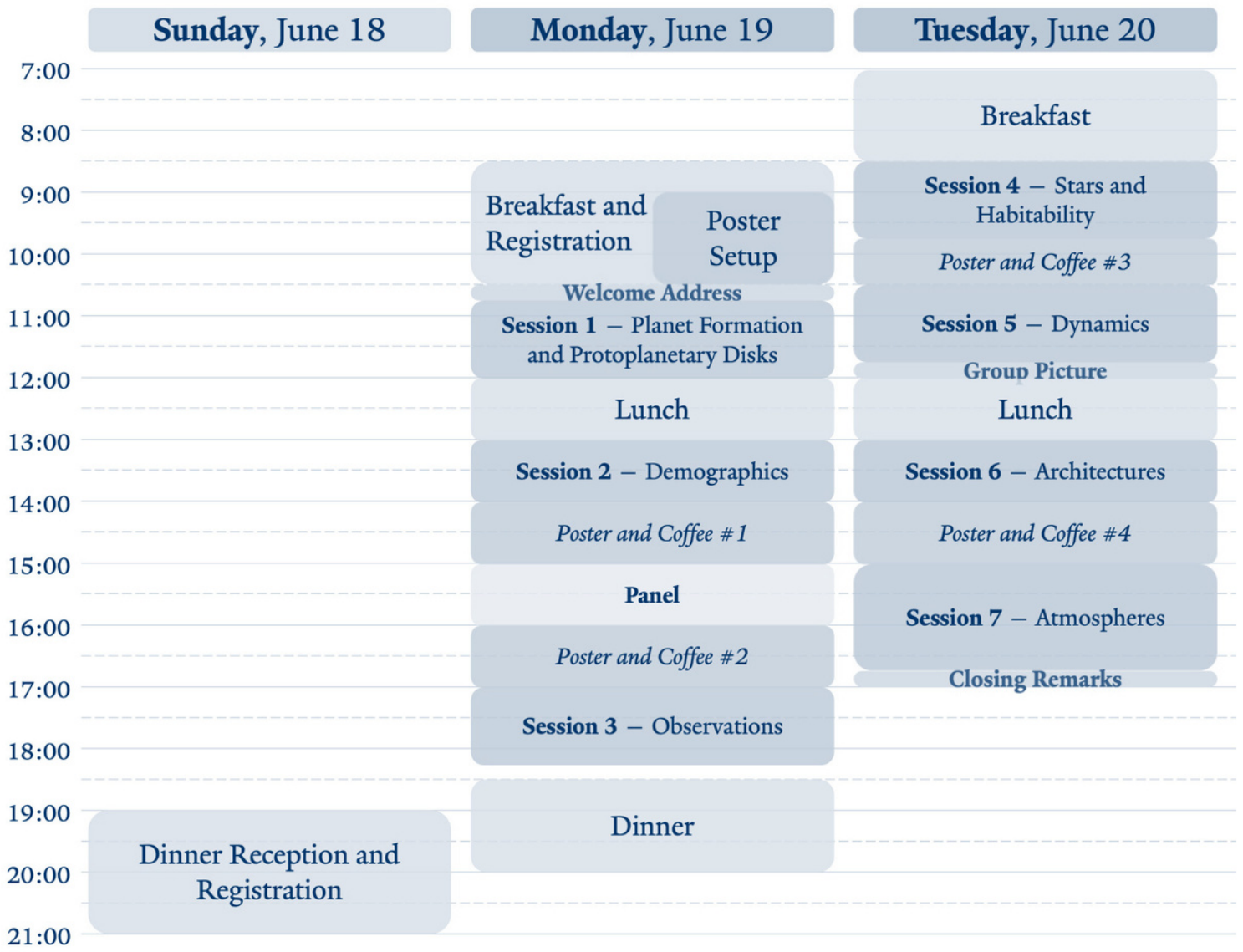}
\caption{Block schedule for ERES 2023. Attendees were also given a PDF abstract book with more detailed information on the scientific program and conference logistics.}
\label{fig:schedule}
\end{figure}

In Figure \ref{fig:schedule}, we show the block schedule that was disseminated to participants in the abstract book. Two events that were added later were the trivia night on Monday, June 19, and the Yale campus tour on Tuesday, June 20.

\section{ERES 2023 Code of Conduct} \label{sec:CoCText}

ERES 2023 is committed to maintaining a professional environment that promotes respect, equality of opportunity, and fair treatment for all individuals. All participants of ERES 2023 at Yale, including the organizing committee and external attendees, are expected to adhere to the following principles of conduct toward others:\\

\textbf{Respect}: All people encountered in a professional setting should be treated with respect at all times. Abusive, demeaning, humiliating, or intimidating behavior is unacceptable and will not be tolerated. This includes behavior based on factors such as gender, race, ethnic and national origin, religion, age, marital status, sexual orientation, gender identity and expression, disability, veteran status, or any other protected class.\\

\textbf{Free Expression and Exchange of Ideas}: ERES 2023 encourages the free expression and exchange of scientific ideas. Scientists should create an environment that fosters open discussion and respectful disagreement, regardless of differences in opinions or positions.\\

\textbf{Unacceptable Behaviors}: ERES 2023 prohibits behaviors that are considered unacceptable, including but not limited to:

\begin{itemize}
    \item \textit{Harassment}: Harassment based on factors such as race, religion, color, gender, age, national origin, disability, marital status, sexual orientation, gender identity and expression, or any other protected class, is strictly prohibited. This includes verbal harassment (e.g., offensive comments, slurs, teasing, stereotyping), nonverbal harassment (e.g., distribution or display of inappropriate written or graphic material), and physical harassment (e.g., physical assault or violating personal space).
    \item \textit{Sexual Harassment}: Sexual harassment, including ``quid pro quo" harassment or creating a hostile work environment, is unacceptable. This includes unwelcome sexual attention, requests for sexual favors, sexually oriented conduct that interferes with job performance, or creating an offensive or unpleasant working environment.
    \item \textit{Bullying}: Bullying behavior, such as demeaning, intimidating, or sabotaging the work of others, whether as individuals or as a group, is unacceptable. This includes verbal bullying (e.g., threatening, slandering, ridiculing), physical bullying (e.g., pushing, assaulting), gesture bullying (e.g., nonverbal threatening gestures), or sabotaging an individual's work.
\end{itemize}

\textbf{Professionalism}: Attendees at ERES 2023 should be expected to maintain a high level of professionalism in their interactions with others. This includes treating all attendees with respect, refraining from engaging in discriminatory, harassing, or abusive behavior, and maintaining a positive and inclusive atmosphere for all participants.\\

\textbf{Responsibility of Senior Members}: Senior attendees, including research supervisors, have a special responsibility to facilitate the research, educational, and professional development of students and subordinates. This includes providing safe and supportive work environments, appropriate acknowledgment of their contributions to research results, and protection of their academic freedom. It also includes supporting the career aspirations of undergraduate students, graduate students and young professionals, whether they choose an academic or non-academic career path.\\

\textbf{Presentation Ethics}: Presenters at ERES 2023 should adhere to ethical guidelines for presenting and publishing their research. This includes properly citing the work of others, avoiding plagiarism, and presenting research findings accurately and transparently. Any potential conflicts of interest should be disclosed, and presenters should refrain from misrepresenting their research or making exaggerated claims.\\

\textbf{Intellectual Property}: Attendees should respect the intellectual property rights of others, including not using or sharing others' data, results, or ideas without proper attribution or permission. Attendees should also be mindful of protecting their own intellectual property, such as research ideas or unpublished findings, and take appropriate measures to prevent unauthorized use or disclosure.\\

\textbf{Responsible Conduct of Research}: Attendees should adhere to responsible research practices, including conducting research in an ethical, transparent, and reproducible manner. This includes following appropriate research methodologies, handling data and materials responsibly, and maintaining integrity in reporting research findings.\\

\textbf{Collaborative and Collegial Conduct}: Attendees should foster a collaborative and collegial environment at the ERES 2023, promoting constructive discussions, exchanging ideas and knowledge, and engaging in respectful debate. Attendees should also respect the diversity of opinions, backgrounds, and perspectives among conference participants and avoid engaging in behavior that may harm professional relationships or create a hostile environment.\\

\textbf{Compliance with Laws and Regulations}: Attendees should comply with all applicable laws, regulations, and policies relevant to the conference, including those related to ethics, data protection, safety, and intellectual property. This includes obtaining necessary permissions, approvals, or permits for conducting research, and adhering to any local or regional regulations governing the conference.\\

\textbf{Slack Etiquette}: This code of conduct also applies to virtual interactions on the associated Slack workspace leading up to, during, and after the conference. Attendees should maintain a respectful and professional environment, use appropriate language, follow channel guidelines, maintain confidentiality, be responsive and collaborative, and report any inappropriate behavior.\\

\textbf{Handling Potential Breaches of the Code of Conduct}: If you experience or witness a breach of the code of conduct, it is important to report it promptly. We take breaches of the code of conduct seriously and will take appropriate action to address the situation. This may result in consequences such as warnings, removal from the conference, or other appropriate actions.

To report an incident, please contact \textit{redacted}, who form the Conduct \& Safety Committee within ERES 2023, either in-person, via email, or on Slack. You can also reach out to any other member of the organizing committee.\\

\textbf{Further Reading}: ERES 2023 is committed to following the guidelines outlined in the AAS Code of Ethics. 

\section{Post-Conference Survey} \label{sec:survey}

\begin{figure}[!h]
\plotone{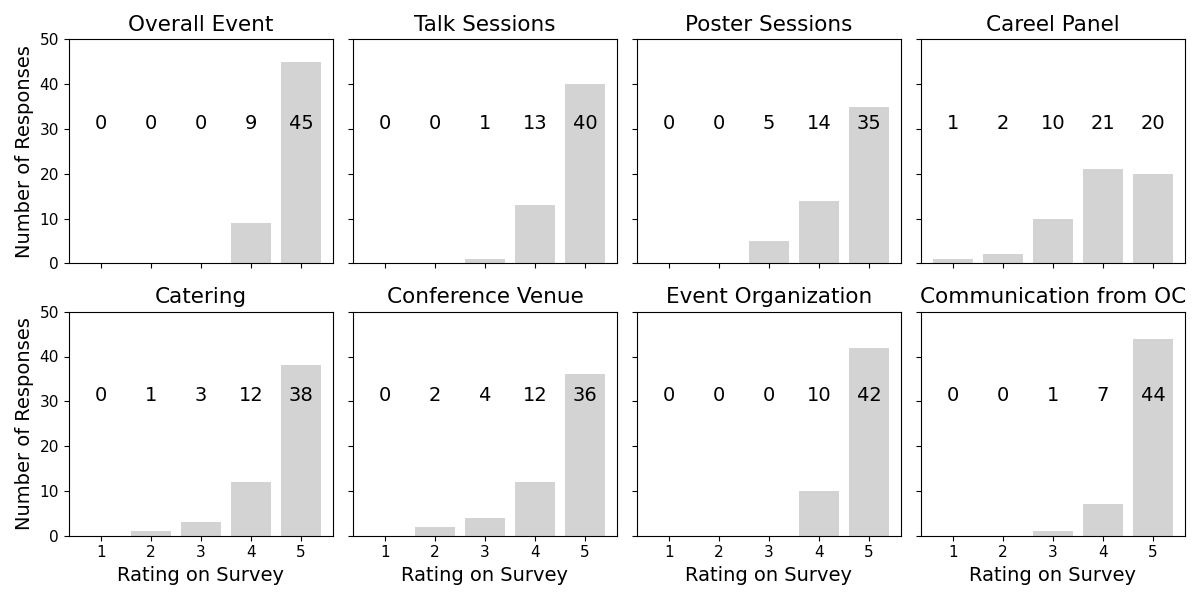}
\caption{Selected results from the optional ERES post-conference survey. All scores were given on a scale where 1 and 5 are the least and most favorable, respectively. Attendees gave favorable ratings to all aspects of the conference and were especially positive on communication from the OC.}
\label{fig:multipanel}
\end{figure}

Here, we present additional results of the post-conference survey introduced in Section \ref{sec:conference}. Although the form was anonymous, we elect not to present any answers from text-box questions for privacy purposes. The breakdown of participants who completed the survey was as follows: 20 graduate students in at least their fourth year, 14 graduate students in years 1-3, 10 postdocs, and 7 undergraduates, amounting to about 40\% of attendees. Figure \ref{fig:multipanel} shows the results from several questions where participants were asked to score various aspects of ERES 2023 on a 1-5 scale. The quantitative results have provided important feedback that will be considered when planning ERES 2024.

The written feedback reflected the overall positive tone about the opportunities for professional development that a conference for early-career researchers provided and a desire for more time in a safe space with peers. There were multiple requests for an additional day of the conference for more presentations and flash talks for poster presenters.

\section{Conference Gantt Chart} \label{sec:Gantt}

\begin{figure}[!h]
\centering
\includegraphics[width=1\linewidth]{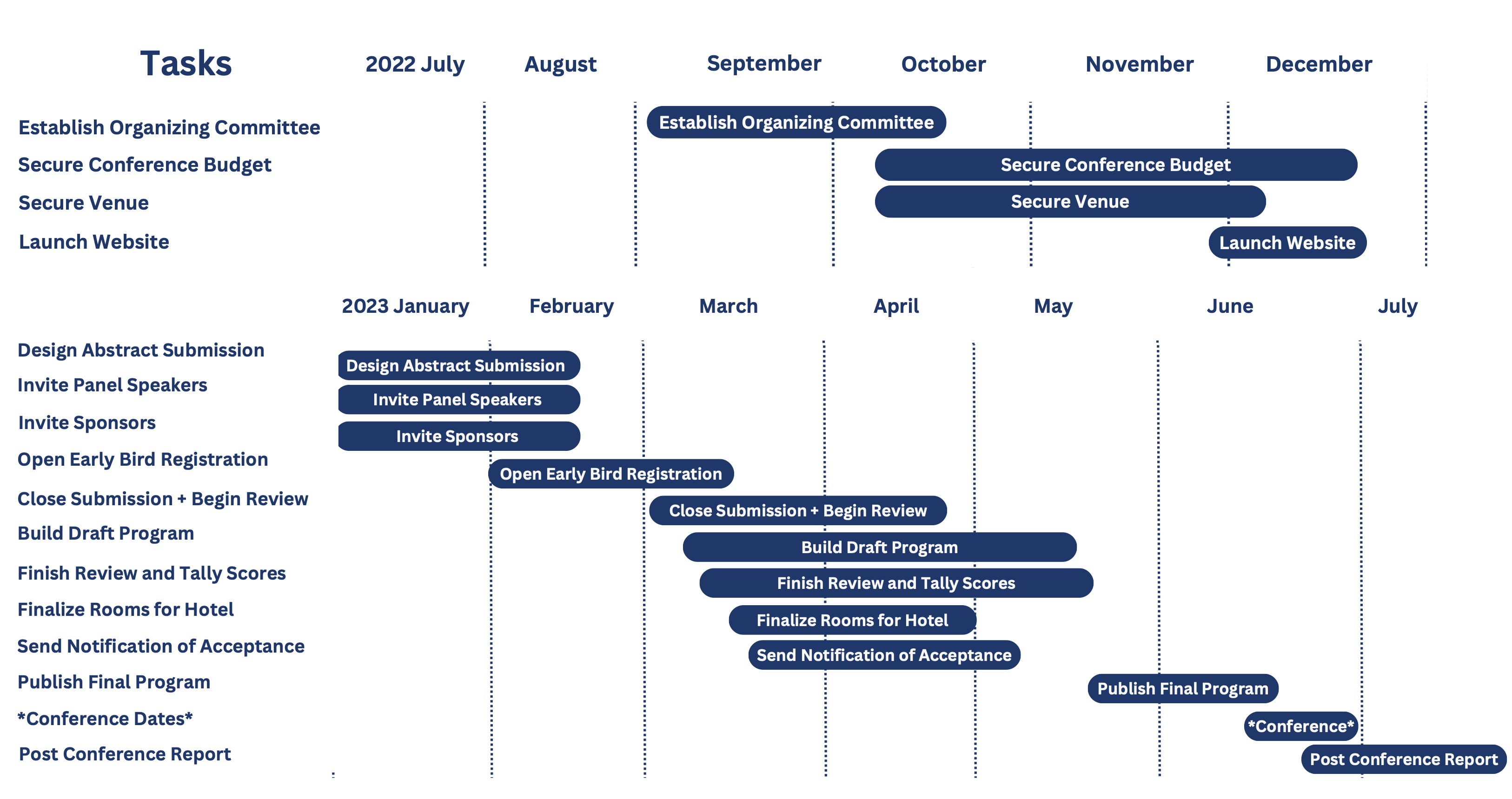}
\caption{Gantt project planning schedule for ERES 2023 that was followed by the OC for conference organization. }
\label{fig:gantt}
\end{figure}

In Figure \ref{fig:gantt}, we show the timeline used to plan key steps in the conference organization process. The OC used Notion to keep a running list of tasks that could be assigned to individuals and viewed as a timeline, to-do list, or table.

We did not use professional software to organize either the logistics or the scientific program. We used Google Docs and Google Sheets to work collaboratively and used Notion to keep a running to-do list. If ERES symposia continue to exceed 100 attendees, then specialized conference organization software might be a worthwhile investment.

\end{document}